\documentclass[conference]{IEEEtran}

\usepackage{cite}
\usepackage{amsmath,amssymb,amsfonts}
\usepackage{graphicx}
\usepackage{booktabs}
\usepackage{libertinus}
\usepackage[colorlinks=true,
            linkcolor=blue,    
            citecolor=blue,    
            urlcolor=blue]     
\usepackage{xcolor}

\usepackage{soul}

\begin{document}

 \title{Quantum-Resistant Networks Using Post-Quantum Cryptography}

\author{
\IEEEauthorblockN{
Xin Jin\IEEEauthorrefmark{3}$^\text{1}$,
Nitish Kumar Chandra\IEEEauthorrefmark{4}$^\text{1}$,
Mohadeseh Azari\IEEEauthorrefmark{4},
Kaushik P. Seshadreesan\IEEEauthorrefmark{4}
Junyu Liu\IEEEauthorrefmark{3},
}
\IEEEauthorblockA{\IEEEauthorrefmark{3}\textit{Department of Computer Science, University of Pittsburgh, Pittsburgh, PA 15260, USA}}
\IEEEauthorblockA{\IEEEauthorrefmark{4}\textit{Department of Informatics and Networked Systems, University of Pittsburgh, Pittsburgh, PA 15260, USA}}

\IEEEauthorblockA{Emails: xij90@pitt.edu, nkc16@pitt.edu, MOA125@pitt.edu, KAUSESH@pitt.edu, junyuliu@pitt.edu}
}

\maketitle

\begin{abstract}

Quantum networks rely on both quantum and classical channels for coordinated operation. Current architectures employ entanglement distribution and key exchange over quantum channels but often assume that classical communication is sufficiently secure. In practice, classical channels protected by traditional cryptography remain vulnerable to quantum adversaries, since large-scale quantum computers could break widely used public-key schemes and reduce the effective security of symmetric cryptography. This perspective presents a quantum-resistant network architecture that secures classical communication with post-quantum cryptographic techniques while supporting entanglement-based communication over quantum channels. Beyond cryptographic protection, the framework incorporates continuous monitoring of both quantum and classical layers, together with orchestration across heterogeneous infrastructures, to ensure end-to-end security. Collectively, these mechanisms provide a pathway toward scalable, robust, and secure quantum networks that remain dependable against both classical and quantum-era threats.
\end{abstract}

\begin{IEEEkeywords}
Quantum Networks, Post-Quantum Cryptography, Quantum Memory, Coherence Time, Entanglement Distribution  \footnote{:These authors contributed equally.}
\end{IEEEkeywords}

\section{Introduction}

Quantum entanglement is a fundamental resource in quantum communication, enabling protocols that achieve information transfer beyond the capabilities of classical systems. Quantum key distribution (QKD)~\cite{doi:10.1126/science.283.5410.2050,BENNETT20147}, quantum teleportation~\cite{PhysRevLett.70.1895}, and entanglement swapping~\cite{davis2025entanglementswappingsystemsquantum} demonstrate how quantum states can be transmitted securely across distance based on the principles such as superposition and the no-cloning theorem~\cite{Wootters1982}. These advances have laid the groundwork for large-scale quantum networks, where entanglement distribution serves as the foundation for secure communication and distributed quantum information processing~\cite{Kimble2008, doi:10.1126/science.aam9288}.

Most quantum applications, in addition to relying on quantum channels, also depend critically on classical channels for exchanging information. Classical communication is required for reconciliation in QKD~\cite{PhysRevApplied.18.044022}, for transmitting syndrome data in quantum error correction~\cite{PhysRevLett.77.793,gottesman2009introductionquantumerrorcorrection}, and for exchanging measurement outcomes in entanglement purification~\cite{PhysRevLett.76.722}. This reliance creates a significant vulnerability: while the quantum components of these protocols may be intrinsically resistant to eavesdropping, the classical components continue to depend on conventional cryptography, which is threatened by quantum adversaries~\cite{8490169}. If attackers possess quantum computational power, classical authentication and coordination mechanisms can be compromised, undermining the very security that quantum applications are designed to provide.

Classical cryptography secures current digital communication, but these protocols are vulnerable in the presence of quantum computers. Shor’s algorithm~\cite{doi:10.1137/S0097539795293172,365700} can solve integer factorization and discrete logarithms in polynomial time, breaking the hardness assumptions underlying RSA~\cite{10.1145/359340.359342}, Diffie Hellman key exchange~\cite{1055638}, the Digital Signature Algorithm (DSA)~\cite{pub2000digital}, and elliptic curve cryptography (ECC)~\cite{koblitz1987elliptic}. In addition, Grover’s algorithm~\cite{10.1145/237814.237866} provides a quadratic speedup for brute force search, reducing the effective security level of symmetric key ciphers and hash functions by half; for instance, AES 128 offers only about 64 bits of security against a quantum adversary~\cite{scrivano2025comparativestudyclassicalpostquantum}. Together, these results show that conventional cryptographic primitives cannot provide long term security in a quantum era.

In response to the looming threat posed by quantum computers, the U.S. National Institute of Standards and Technology (NIST) initiated its Post-Quantum Cryptography (PQC) standardization project in 2016~\cite{chen2016report}. After several evaluation rounds, NIST announced in July 2022 the first algorithms selected for standardization: CRYSTALS–Kyber as a key encapsulation mechanism and CRYSTALS–Dilithium as a digital signature scheme~\cite{8406610,cryptoeprint:2017/633}, together with SPHINCS+~\cite{hulsing2019sphincs+}, a stateless hash-based signature system. These algorithms were formally adopted as Federal Information Processing Standards (FIPS) in 2024~\cite{10984484}. The selections reflect the current consensus that lattice-based cryptography offers the most practical combination of efficiency and security, while SPHINCS+ was included to ensure diversification beyond lattice-based assumptions.

Although quantum protocols are designed to provide information-theoretic security, practical implementations have revealed exploitable weaknesses. Side-channel attacks such as detector efficiency mismatch~\cite{PhysRevA.74.022313}, time-shift strategies~\cite{PhysRevA.78.042333}, and bright-illumination attacks~\cite{Lydersen2010} demonstrate that adversaries can manipulate devices without triggering disturbance detection. Fei et al.~\cite{Fei2018} showed that man-in-the-middle attacks on calibration phase vulnerabilities can enable basis dependent efficiency mismatches that leak key information. Source side vulnerabilities have also been investigated: Trojan horse probing of transmitters allows adversaries to inject light and retrieve internal settings, while injection locking attacks can force lasers to operate under the attacker’s influence~\cite{Jain_2014,PhysRevX.5.031030,PhysRevApplied.13.034008}. Together, these findings show that quantum protocols cannot be regarded as intrinsically secure in practice, as weaknesses in physical components and other real world constraints create exploitable vulnerabilities.

 Various countermeasures against quantum side channel and device level attacks have been proposed, such as measurement device independent QKD and decoy state methods~\cite{PhysRevA.72.012326,PhysRevLett.108.130502,PhysRevLett.108.130503}, but these remain an active area of research, and new approaches continue to be developed. In contrast, the protection of the classical channels that underpin the quantum communication protocols has received comparatively less attention, even though they represent a critical vulnerability. Ensuring the authenticity and confidentiality of reconciliation messages and coordination signals requires solutions capable of withstanding adversaries equipped with quantum computational capabilities. The security of the classical layer in the quantum era remains a missing link in current designs, and addressing this gap through post quantum cryptographic protocols is essential. The main challenges in deploying these protocols include achieving interoperability with existing infrastructure, balancing efficiency tradeoffs, and ensuring seamless integration with quantum specific operations, all of which are necessary steps toward building fully robust quantum networks~\cite{10602299,10.1109/COMST.2023.3294240}.

 \section{Quantum Resistant Networks}

\begin{figure}[h]
    \centering
    \includegraphics[width=0.35\textwidth]{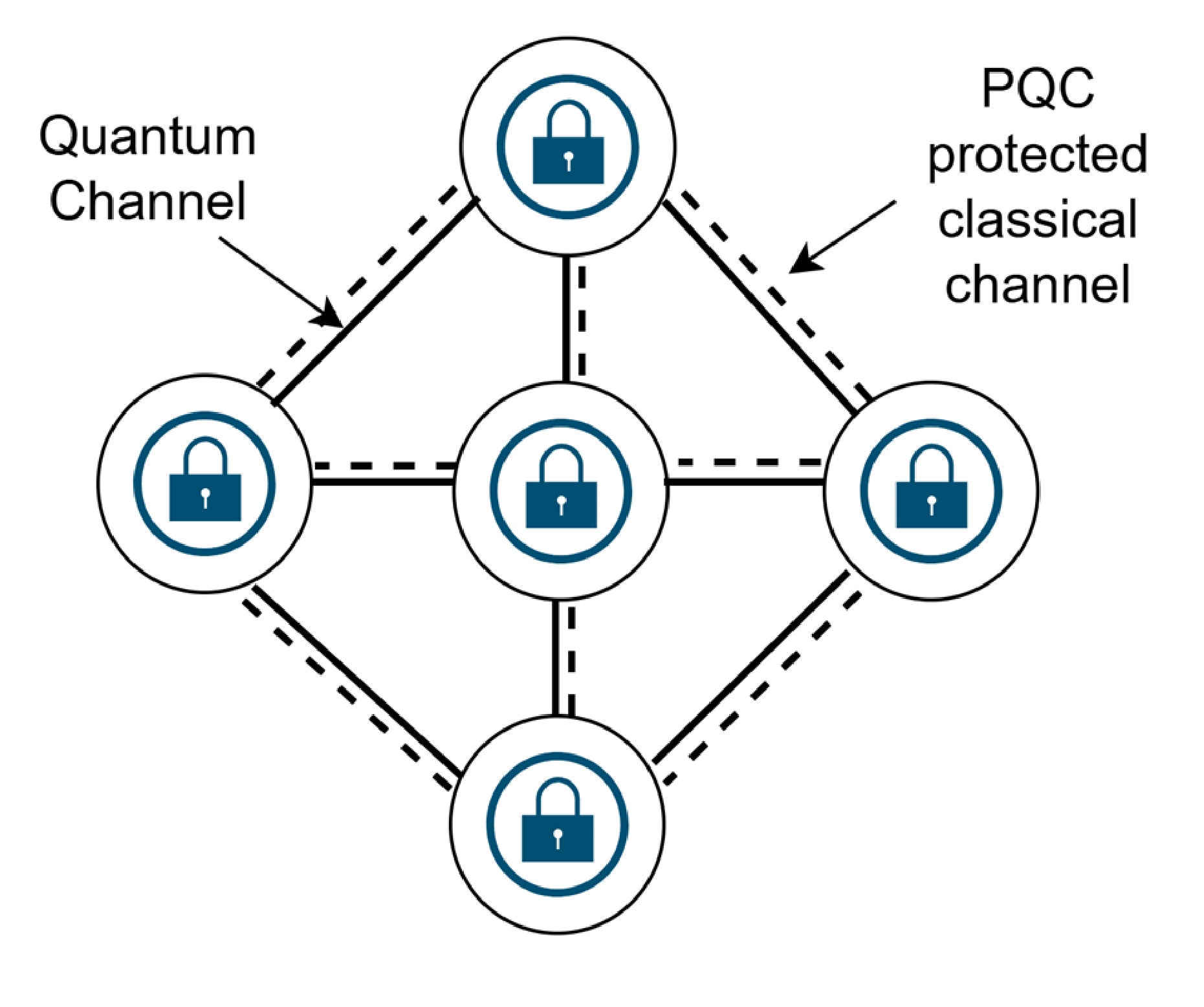}
    \caption{Schematic of a quantum network represented as nodes connected by edges. 
    Each edge consists of a quantum channel (solid line) and a PQC-protected classical channel (dotted line), 
    ensuring secure communication between nodes.}
    \label{quantum_network}
\end{figure}

A quantum network consists of nodes such as quantum processors, repeaters, or end user devices interconnected by quantum channels that distribute entanglement across the network (See Fig.~\ref{quantum_network}). These quantum links enable protocols such as entanglement swapping and quantum teleportation. While quantum channels carry qubits and entanglement, the successful execution of network operations critically depends on classical communication. For example, in teleportation or entanglement swapping, the outcomes of Bell state measurements must be transmitted to remote nodes so that corrective operations can be applied~\cite{doi:10.1126/science.aam9288}. Similarly, classical channels are indispensable for synchronization and control signaling across the network.  

Protecting this classical layer against adversaries equipped with quantum computers requires embedding post quantum cryptography (PQC) into each stage of the protocol stack. Without PQC, authentication and confidentiality of measurement outcomes, synchronization data, or routing messages would be vulnerable to interception or manipulation, undermining the security guarantees of the quantum layer itself. The challenge, however, is that PQC introduces computational overhead for encryption and decryption in addition to the intrinsic latency of classical message exchange. Because qubits must be stored in quantum memories while awaiting these classical signals, the timing of classical communication becomes tightly coupled to the coherence limits of available memory technologies.  

\subsection*{{Timing Constraints for PQC Protected Communication}}

In quantum networks, qubits are often stored in quantum memories while waiting for classical information needed to complete distributed operations such as teleportation or entanglement swapping. If the classical message arrives too late, the stored quantum state decoheres and the protocol fails. When post quantum cryptography (PQC) is used to protect classical communication, additional delay is introduced by cryptographic operations. To ensure correct functioning, the total delay must remain below the memory coherence time.

\paragraph{Notation used in following equations}  
\begin{itemize}
  \item $T_{\text{encrypt}}$: time required to encrypt or authenticate the classical message before transmission.  
  \item $T_{\text{comm}}$: classical communication delay, including propagation through the channel and processing in the network.  
  \item $T_{\text{decrypt}}$: time required to decrypt or verify the received message.  
  \item $T_{\text{coh}}$: coherence time of the quantum memory storing the qubit during the wait.  
  \item $i$: index of a repeater or round of communication.  
  \item $\mathcal{P}$: set of repeaters whose Bell state measurement (BSM) results are sent to the end node.  
  \item $L$: number of dependent rounds of classical communication in a sequential protocol.  
\end{itemize}

\paragraph{Single-hop case}  
For a single sender–receiver pair, the qubit in memory must survive the combined time for encryption, transmission, and decryption:
\[
T_{\text{encrypt}} + T_{\text{comm}} + T_{\text{decrypt}} < T_{\text{coh}}.
\]
This form applies to simple two-node protocols such as quantum teleportation.

\paragraph{Multi-hop with parallel communication}  
In repeater-based networks, many BSMs may be performed simultaneously at different nodes, and their outcomes are broadcast in parallel. The end node must wait until the \emph{last} relevant message arrives. The condition becomes
\[
\max_{i \in \mathcal{P}} \big( T_{\text{encrypt},i} + T_{\text{comm},i \to \text{end}} + T_{\text{decrypt},\text{end}} \big) 
< T_{\text{coh}}^{(\text{end})}.
\]
Here, the maximum determines the constraint because all messages are sent simultaneously, and the slowest one sets the waiting time.

\paragraph{Protocols with sequential signaling}  
Some protocols, such as certain entanglement purification or multi-round handshake schemes, may require multiple dependent rounds of classical communication. In such cases, the delays accumulate:
\[
\sum_{i=1}^{L} \big( T_{\text{encrypt},i} + T_{\text{comm},i} + T_{\text{decrypt},i} \big) 
< T_{\text{coh}}.
\]
This cumulative bound applies whenever each round must be completed before the next begins.

\medskip
In summary, the appropriate inequality depends on the protocol structure: the \emph{single-hop} case for one-shot exchanges, the \emph{parallel} case when multiple classical information are broadcast simultaneously, and the \emph{sequential} case when classical messages must be exchanged in a round-by-round fashion.

Designing PQC integration strategies that respect these timing constraints requires careful attention to cryptographic overheads. Practical approaches include minimizing encryption and verification latency, exploiting parallelism where protocol structure allows, and pre-establishing secure keys so that cryptographic operations do not lie on the critical path of quantum communication. Aligning PQC operations with quantum state preparation and memory usage is essential to ensure that the total classical delay remains within the coherence window. Addressing these challenges is necessary for quantum networks to be resilient against adversaries equipped with quantum computational capabilities.

\subsection*{{PQC algorithm selection for heterogeneous network nodes}}

A single PQC algorithm will not be suitable for all scenarios in a large scale quantum resistant network. Different types of network nodes such as resource constrained quantum processors, high throughput quantum switches, and routers operating over satellite or fiber based communication channels require different PQC choices depending on their computational power, link latency, memory limits, and required security level. Edge nodes such as user terminals may prefer lightweight algorithms (for example, lattice based KEMs with smaller key sizes such as Kyber512) that minimize encryption and decryption times. In contrast, core nodes or quantum repeaters, which have greater processing capacity and stricter security requirements, can afford more computationally intensive algorithms such as FrodoKEM 1344. Selecting PQC algorithms therefore requires balancing key size, computational load, and expected attack complexity to match the capabilities and roles of heterogeneous devices in the network~\cite{unsal2025comparativeperformanceevaluationkyber}.

\subsection*{{Quantum Memory Hierarchy and Architectural Adaptations}}
A hierarchy of quantum memories is critical for integrating PQC while preserving overall network performance. Just as classical computing employs a tiered memory structure (cache, RAM, disk) to balance speed and capacity, quantum networks can utilize memories with different coherence times and access speeds at different layers. Long lived quantum memories, such as those based on trapped ions or error corrected logical qubits, are best suited for backbone nodes where repeaters must store entangled states while awaiting classical feedforward messages from neighboring links. In contrast, short lived quantum memories, such as photonic or atomic ensemble memories, can be used to buffer rapidly generated local entanglement between adjacent nodes, which is swapped forward almost immediately once both links succeed. This layered approach allows PQC protected classical communication to be processed within coherence constraints by aligning memory lifetimes with the expected classical delays at each level of the network~\cite{Heshami12112016}.

\section{Man-In-The-Middle for hybrid quantum-classical network}
\subsection*{{Hybrid Quantum-Classical Adversary Model}}  
 
We consider a hybrid adversary capable of exploiting both the quantum and classical layers of the network. On the quantum side, the adversary may intercept qubits in transit and load them into a quantum memory, which requires a finite interception and storage latency denoted by $T_{\text{Eve}}$. The stored qubits can only be preserved for the coherence time of the adversary’s memory, $T_{\text{coh}}^{\text{Eve}}$. On the classical side, the adversary may manipulate coordination messages, such as teleportation corrections or entanglement swapping, by spoofing, delaying, or relaying PQC-protected communication. The additional delay introduced by these manipulations is denoted by $T_{\text{pqc}}$. The total adversarial delay is therefore
\[
\Delta t = T_{\text{Eve}} + T_{\text{pqc}}.
\]
A man-in-the-middle attack can only succeed if the adversary is able to complete both the quantum interception and the classical manipulation before decoherence occurs, that is if
\[
\Delta t < T_{\text{coh}}^{\text{Eve}}.
\]

If this bound is exceeded, the adversary’s stored quantum states undergo decoherence, resulting in increased quantum bit error rates (QBER) or reduced entanglement fidelity, thereby making the intrusion detectable. By explicitly accounting for finite coherence times and PQC-induced delays, this model moves beyond earlier idealized assumptions and establishes a framework for analyzing realistic joint quantum–classical attack vectors in quantum networks.

\subsection*{{Mitigation Strategies for Hybrid MITM Attacks}}

Robust defense against hybrid man-in-the-middle adversaries requires coordinated measures across both the quantum and classical layers~\cite{RevModPhys.92.025002}. All classical coordination traffic should be protected with post-quantum cryptographic authentication to prevent spoofing, replay, etc. To move beyond static threshold tests, anomaly detection techniques such as machine learning models trained on expected error patterns, fidelity distributions, timing statistics, etc., can provide early warning of subtle intrusions that might otherwise remain hidden. At the network level, robustness can be further enhanced through multipath routing of both quantum states and classical information, which forces an adversary to compromise multiple channels simultaneously. Taken together, these measures create a layered defense framework that significantly increases the cost and complexity of sustaining hybrid quantum classical attacks.

\section{Towards Securing Large Scale Quantum Networks}

\subsection*{{PQC-Orchestrated Key Infrastructure at Scale}}

A robust Key Management System (KMS) is required to orchestrate network-wide key establishment and frequent re-keying, scaling from point-to-point links to multi-node topologies~\cite{10.1007/978-3-030-75539-3_7}. In a fully connected network, each node must exchange new keys with every other node, leading to $H(N) \sim \mathcal{O}(N^2)$ handshakes per re-key cycle. To control this growth, hierarchical or orchestrated key infrastructures can be employed, reducing complexity toward near-linear scaling. The design objective is to minimize $T_{\text{key}}$ (the time for key rotation) while keeping $T_{\text{auth}}$ low and $H(N)$ tractable as $N$ grows. Achieving this balance ensures that the classical coordination layer of the quantum network remains quantum resistant without imposing prohibitive latency or overhead.

\subsection*{{Physical Constraints to Secure Quantum Networks}}

Maintaining high-quality entanglement across long distances and heterogeneous links (optical fiber, free space, and satellite based channels) is a central challenge for large-scale quantum networks. Routing protocols must incorporate entanglement swapping at intermediate nodes together with techniques such as entanglement purification or local error correction~\cite{Yan2023}. Two key constraints are the quantum memory coherence time $T_{\text{coh}}$ (the maximum time a stored qubit remains coherent) and the entanglement generation rate $R_e$ (the rate of producing entangled pairs per link). For secure operation, the total end-to-end entanglement distribution time $T_{\text{dist}}$ must remain below the coherence time, i.e., $T_{\text{dist}} < T_{\text{coh}}$, ensuring that stored pairs are still usable when multi-hop processes are completed. Synchronization across hops is also necessary so that multiple links generate entanglement within a common window, effectively requiring $R_e T_{\text{coh}} \gg 1$, which guarantees that many attempts can succeed within one memory lifetime and reduces adversarial opportunities to exploit timing gaps.

Another critical factor is the fidelity of the quantum state. The end-to-end fidelity $F_{\text{end}}$ of an $L$-hop entangled link decreases with each swap operation and is strongly constrained by the lowest-quality link or memory in the chain. Transmission losses, imperfect swapping, and memory decoherence degrade quantum states with time, and without active entanglement distillation they can quickly push $F_{\text{end}}$ below application thresholds. From a security standpoint, low fidelity not only limits performance but also obscures the distinction between natural noise and malicious interference, making adversarial actions harder to detect. Ensuring that fidelity remains above threshold is therefore essential for both reliable and secure operation of large-scale quantum networks.

\section{Conclusion}
This perspective discusses how achieving fully quantum-resistant networks requires moving beyond treating cryptography, entanglement distribution, and network control as separate components. Future systems must integrate these elements into a unified architecture capable of sustaining end-to-end security under realistic timing and adversarial constraints. We have discussed how post-quantum cryptographic techniques can be embedded into protocol lifecycles, how a hybrid adversary model exposes vulnerabilities overlooked by traditional approaches, and how scalable routing together with machine learning methods can strengthen reliability. A fully quantum-resistant network will also depend on advances in quantum memory technologies, efficient PQC implementations, and coordinated control frameworks spanning heterogeneous infrastructures.

At the same time, significant research challenges remain. Open problems include scaling deployment to ultra-long distances, maintaining robustness under high levels of noise, developing routing algorithms for complex and dynamic topologies, and mitigating failure scenarios such as repeater congestion under multi-user access. Addressing these issues is critical before quantum-resistant networking can move from conceptual proposals to practical, global-scale systems.

\section*{Acknowledgments}
JL is supported in part by the University of Pittsburgh, School of Computing and Information, Department of Computer Science, Pitt Cyber, Pitt Momentum, PQI Community Collaboration Awards, John C. Mascaro Faculty Scholar in Sustainability, NASA under award number 80NSSC25M7057, and Fluor Marine Propulsion LLC (U.S. Naval Nuclear Laboratory) under award number 140449-R08. This research used resources of the Oak Ridge Leadership Computing Facility, which is a DOE Office of Science User Facility supported under Contract DE-AC05-00OR22725. 

\bibliographystyle{IEEEtran}
\bibliography{new_ref} 
\end{document}